# All-dielectric hybrid VIS-NIR dual-function metasurface

PEI XIONG,[1] DANIEL K. NIKOLOV,[1] FEI CHENG,[1] JANNICK P. ROLLAND[1,4] AND A. N. VAMIVAKAS[1,2,3,5,*]

[1]*Institute of Optics, University of Rochester, 480 Intercampus Dr, Rochester, NY 14627, USA*
[2]*Department of Physics, University of Rochester, 500 Wilson Blvd, Rochester, NY 14627, USA*
[3]*Materials Science, University of Rochester, Rochester, NY 14627, USA*
[4]*Center for Freeform Optics, University of Rochester, Rochester, NY 14627, USA*
[5]*Center for Coherence and Quantum Optics, University of Rochester, Rochester, NY 14627, USA*
*nick.vamivakas@rochester.edu

**Abstract:** Metasurfaces are a promising technology that can serve as a compact alternative to conventional optics while providing multiple functions depending on the properties of the incident light, such as the wavelength, polarization, and incident angle. Here, we demonstrate a hybrid VIS-NIR dielectriac metasurface that can reflect 940 nm light into a specified direction while transmitting visible light (450-750 nm). The dual functionality is realized by combining an aperiodic distributed Bragg reflector with dielectric meta-tokens. Experimental demonstration is also reported, showing an anomalous reflection of near-infrared (NIR) light within a 20º full field-of-view (FOV) and the transmission of wavelengths from 450 nm to 750 nm.

## 1. Introduction

Metasurfaces are planar optical elements composed of artificially fabricated nanostructures with tailored optical responses. Recently, they have attracted considerable interest owing to their potential control of the amplitude, phase, and polarization of light [1-4]. This manipulation of the properties of light for metasurfaces has demonstrated a wide range of applications, including wavefront control [5, 6], dispersion compensation [7, 8], and metalens imaging [9-11]. Their flat format is critically well-tailored to the development of miniature optical and electronic devices [12, 13] targeting precise functionalities. As such, it has been shown that metasurfaces can offer alternatives to traditional optical elements, like some lenses [14, 15], gratings [16, 17], polarizers [18, 19], and holograms [20, 21].

Metasurfaces, when combined with dielectric Bragg reflectors (DBRs), can offer additional functionality and control over the manipulation of light [16, 22, 23]. DBRs consist of alternating layers of materials with different refractive indices and they are commonly used in photonics due to their high reflectivity and ability to reflect light at a specific wavelength while transmitting others, making them useful for various applications such as laser mirrors [24] and optical filters [25]. Here, we propose the combination of a metasurface and an aperiodic DBR to develop an all-dielectric hybrid VIS-NIR metasurface that has different control over the NIR and visible wavelength.

Figure 1 shows the schematic diagram of our designed dual-function metasurface. It comprises periodic meta-atoms on top of a thin film substrate. The device can anomalously reflect NIR light within a full FOV of 20º while allowing visible light to transmit directly through. The transmission of visible light is enabled by working with dielectric materials. The direction of the reflected NIR light depends on the period of the top meta-atoms. We co-optimized the thicknesses of selected DBR layers and the geometry of the meta-atoms to simultaneously enhance the reflection efficiency of NIR light [16] and introduce more flexibility to optimize this metasurface for dual functionality. The combination of this aperiodic thin film substrate and the meta-atoms offers a promising avenue for the development of novel optical devices with enhanced performance and functionality.

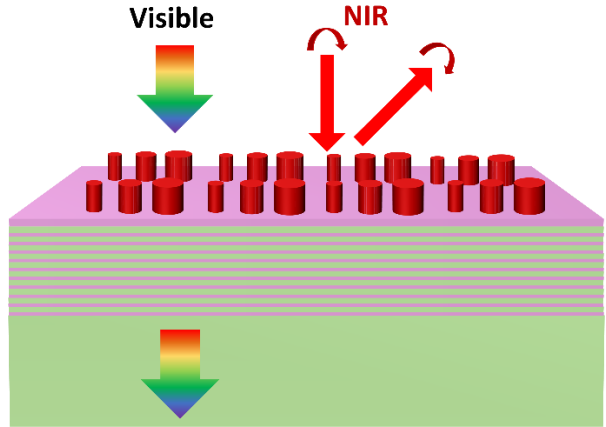

Figure 1. Illustration of the proposed device. It is composed of a $Si_3N_4/SiO_2$ thin-film substrate and a $TiO_2$ nanopillar array that conspire to reflect NIR light in a specific direction while also allowing for the transmission of visible light.

## 2. Simulations

All simulations are done using the rigorous coupled-wave analysis method in RSoft (Synopsys). The DBR substrate consists of silicon nitride ($Si_3N_4$) and silicon dioxide ($SiO_2$) as the high and low index material, respectively. The meta-token material is titanium dioxide ($TiO_2$) due to its high refractive index and low loss. We started with a periodic DBR that has the thickness of each layer fixed to a quarter wave thickness. The central wavelength of the reflectance peak was chosen to be 940 nm, a common wavelength for NIR illuminators. The thickness of each $Si_3N_4$ (n=1.96 @ 940 nm) layer is 120 nm, and each $SiO_2$ (n=1.46 @ 940 nm) layer is 160 nm. This gave us the reflectance spectrum with a flat reflectance around 940 nm and less than 20 % reflectance within the visible spectrum (450 nm – 750 nm) (see Appendix A).

To design the periodic meta-tokens on top of the DBR substrate, we used a 1330 nm by 400 nm unit cell. The pitch size along the *y*-axis was chosen to be less than 450 nm to avoid the possible diffraction of visible light along the direction that is vertical to the grating vector, which we set to be along the *x* direction. The pitch size along the grating vector direction was determined by the grating equation to be 1330 nm assuming 1st order diffraction in air at a diffraction angle of 45° for normally incident light.

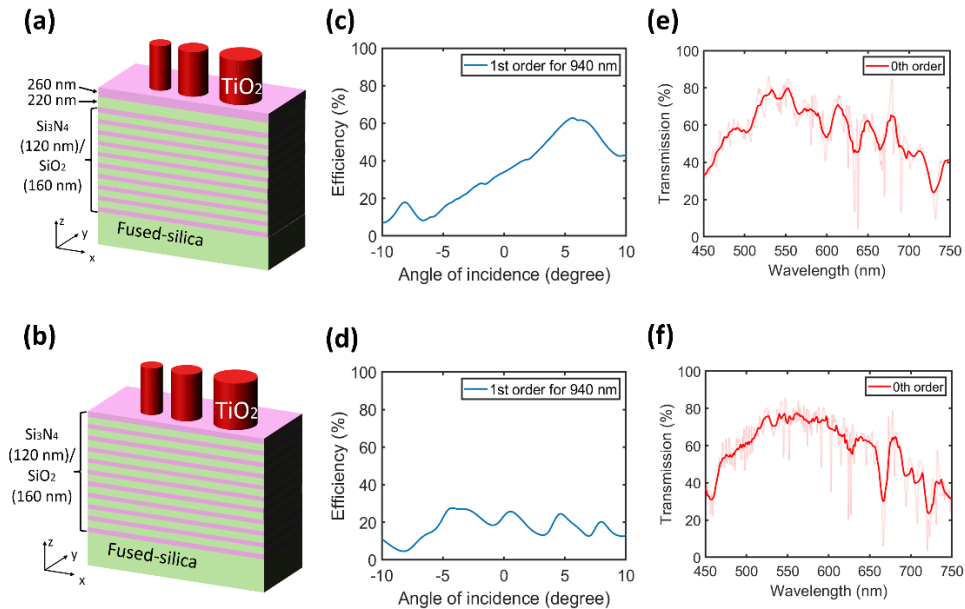

Figure 2. Design of a hybrid VIS-NIR dual-function metasurface: (a) The unit cell for the final optimized metasurface combined with the DBR substrate. The height of the pillars is 540 nm. The diameters for the three pillars are 130 nm, 290 nm, and 350 nm. The center-to-center distance between the smallest pillar and the medium pillar is 260 nm, and between the medium-sized pillar and the largest pillar is 480 nm. The DBR has 21 layers; the optimized thickness of the first layer ($Si_3N_4$)

is 260 nm, and the optimized thickness of the second layer (SiO$_2$) is 220 nm. The rest of the periodic part of the DBR substrate has a thickness of 120 nm for Si$_3$N$_4$ layers and 160 nm for SiO$_2$ layers; (b) The unit cell combining the meta-atoms layer of the optimized design in (a) and the basic periodic DBR substrate; (c) & (d) Simulated 1$^{st}$ order diffraction efficiency as a function of angle of incidence with 940 nm TE polarized light illumination for the metasurface in (a) & (b), respectively; (e) & (f) Simulated transmission spectrum of the metasurface in (a) & (b), respectively. The semi-transparent line is the raw data from the simulation, and the solid red line is the smoothed one using the *smoothdata* function in MATLAB.

Three nanopillars were used in one unit cell to build the metasurface gradient phase. A parameter sweep of the radius and height for a single TiO$_2$ nano pillar placed on top of the DBR substrate was first carried out (see Appendix B). From the sweep results, the reflection amplitude of the single pillar remains near 1 for different radii and heights. When the pillar has a height larger than 400 nm, its reflection phase varies in a range larger than $2\pi$ with different radii, which is necessary to achieve a linear grating phase ramp over the unit cell. The starting point design of the metasurface was first decided based on the parameter sweep results. After that, a general optimization was established using the multi-variable optimization and scanning tool in RSoft. The optimization variables included radii, heights, locations of the three pillars in a unit cell, and the thicknesses of the first two layers of the basic DBR. As discussed in Ref. [16], the thicknesses of the first several layers of the thin-film substrate determine the reflection phase, which combines with the phase variance caused by the meta-tokens, to help increase the anomalous reflection efficiency. So, we allowed the thickness of the first two layers of the DBR substrate to vary to increase the anomalous reflection efficiency of NIR light while maintaining a good transmission of visible light. To optimize the metasurface towards this dual functionality, we used one TE polarized source with a wavelength ranging from 400 nm to 940 nm illuminating from the meta token's side at the normal incident angle. For each simulation, we placed a detector to measure the 1$^{st}$-order diffraction efficiency on the reflection/meta-token side and another one to measure the 0$^{th}$-order diffraction efficiency on the transmission/substrate side. When building the merit function, we aim to maximize the 1$^{st}$-order diffraction efficiency on the reflection side for 940 nm and the average 0$^{th}$-order transmission for the incident wavelength ranging from 450 nm to 750 nm using the same weights for each.

The optimized design parameters are listed in the caption of Fig. 2(a). We evaluated the angular bandwidth of this optimized design for diffracting NIR light by varying the incident angle from -10º to 10º. Referencing the result in Fig. 2(c), over the full FOV of 20º, the average diffraction efficiency is about 40% for this optimized design with a peak of 60% at 5º angle of incidence. Figure 2(e) shows the transmission of the optimized device under an on-axis illumination, and it is about 50% on average for the full visible spectrum. The reason for the lower transmission of the metasurface is the diffraction of visible light during its transmission through the metasurface because of the large grating period (1330 nm) compared to the visible wavelengths (450 nm – 750 nm). We also evaluated the transmission angular bandwidth for this optimized design and the results are depicted in Appendix C.

To demonstrate the effectiveness of modifying the thicknesses of the first two layers of the DBR substrate, we simulated the metasurfaces with the optimized meta-atoms in the above design and the basic periodic DBR substrate, as illustrated in Fig. 2(b). Figure 2(d) displays a maximum efficiency of only 30% for this geometry and it is decreased significantly compared to the device with the aperiodic DBR. The average diffraction efficiency within the angular bandwidth (full FOV of 20º) is around 20%. We also compared the transmission efficiency of these two designs (see Fig. 2(f)) showing that there's no significant change to the 0$^{th}$ order transmission efficiency. This matches well with our expectation that the additional control of the reflection phase achieved by the thickness variation of the top two DBR layers enables a higher reflection efficiency in the NIR wavelength.

**Fabrication**

Figure 3(a) illustrates the metasurface fabrication process. In the first step, a Si$_3$N$_4$/SiO$_2$ DBR mirror was epitaxially grown on a fused silica wafer (University Wafer) using plasmon-enhanced chemical vapor deposition (PECVD). In the second step, the undiluted ZEP-520A (Zeon Chemicals), a positive-tone electron beam resist, was spin-coated on the DBR at the speed of 1800 rpm for 90 seconds to achieve the desired resist thickness of 600 nm, which was set to be larger than the TiO$_2$ nanopillar's height for over-etching in a later step. The resist was then baked at 170 ºC for 2 min. The DisCharge H2Ox2 (DisChem) was spin-coated on the resist at the speed of 2000 rpm for 60 seconds to avoid charging effects during exposure. The patterns were exposed using an accelerating voltage of 100 kV (JEOL 9500). After exposure, the samples were first washed with DI water for 30 seconds to remove the DisCharge H2Ox2. It was developed in ZED-N50 for 3 minutes, after which the sample was transferred to MIBK for 2 minutes to stop the developing process, then rinsed by IPA for 30 seconds and finally blown dry by N$_2$ gas. In the third step, the TiO$_2$ film was coated on the sample using the atomic layer deposition (ALD, Cambridge Savannah 200). A standard two-pulse system of water and the TDMAT precursor was used for running cycles, and each cycle had a 0.015 second water pulse followed by a 25 second delay and a 0.7 second TDMAT pulse followed by the same 25 second delay. The deposition rate is about 0.06 nm per cycle. The chamber was maintained at 100 ºC and the precursor was heated and maintained at 70 ºC throughout the process. In the fourth step, reactive ion etching was carried out on inductively coupled plasma (ICP) reactive ion etching (Oxford Instrument) with a mixture of CHF$_3$ and O$_2$ gas (52 and 2 standard cubic centimeters per minute respectively) at an RF power of 15 W, and ICP power of 2500 W. The etch rate was about 30 nm/min. Finally, in the last step, the sample was cleaned under the oxygen plasma to remove the degraded resist, and the rest of the resist was removed using methylene chloride. Figure 3(b) shows the sample top view under a scanning electron microscope (SEM). The inset is the zoomed-in image with a slanted view. Nanopillars with three different sizes are periodically arrayed in the picture. Figure 3(c) shows the cross-section of the device, which shows that the first two layers of the substrate have different thicknesses from the rest of the layers.

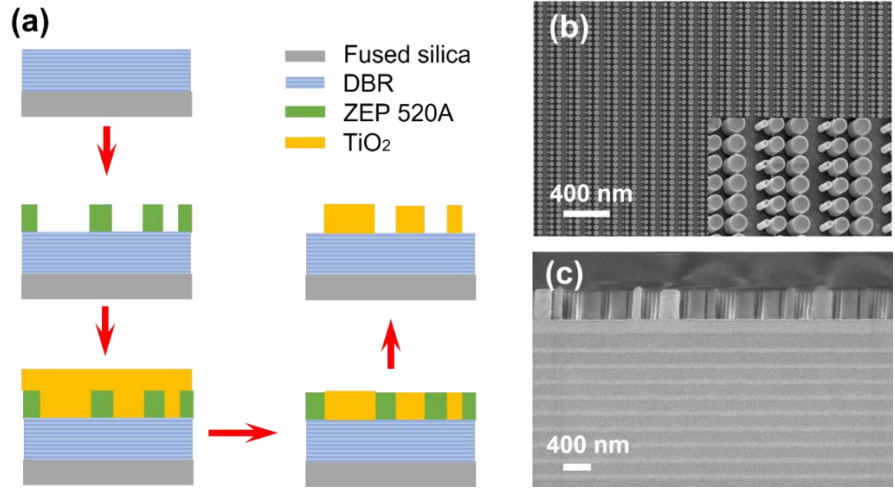

Figure 3. Metasurface fabrication: (a) Fabrication steps for the metasurface device: DBR film is first coated on the fused silica substrate using PECVD, and the inverse pattern is generated on the ZEP 520A resist using electron beam lithography. The TiO$_2$ thin film is coated on the resist pattern conformally using ALD. The ICP-RIE is used to etch the top layer of TiO2 and expose the resist and finally, the resist is removed by both O$_2$ plasma and solvent (b) The top-down SEM image of the fabricated device; the inset is the zoomed-in image at a slanted view. (c) The cross-section image of the fabricated device.

## 3. Testing and discussion

The diffraction efficiency of the fabricated metasurface for near IR light was measured using the setup presented in Fig. 4(a). A Fianium WhiteLase micro supercontinuum laser was combined with a 940 nm, 10 nm full-width-half-maximum (FWHM) bandpass filter as a light source. A linear polarizer filtered the desired TE polarization for the metasurface. The sample was mounted on a custom-built stage, which combines three translation stages along $x$, $y$, and $z$ directions and a rotation stage to provide all degrees of freedom required to align the sample. We also have two other rotation stages aligned center-to-center [circular plate in Fig. 4(a)]. The custom-built stage with the sample was mounted on the arm of one of the two rotation stages for controlling the angle of incidence. The efficiency was measured for angles of incidence from -10° to 10°. On the other arm of the two rotation stages, a lens was mounted to collect the light from the desired diffraction order and to focus it on a Si photodiode.

To better match the plane wave illumination used in simulations, the sample that has the size of 1 mm by 1 mm, was overfilled by the Gaussian profile illumination beam. A knife edge experiment measured the Gaussian beam profile, which was then used to calculate the ratio of the sample area compared with the beam size, which defines the fill factor. The amount of power striking the metasurface could be found by multiplying the total input power by the fill factor. The diffraction efficiency was calculated using the intensity collected by the first Si photodiode divided by the input power on the sample.

The measured first-order diffraction efficiency within 20° full FOV for the fabricated metasurface is shown in Fig. 4(b). The observed efficiency increases from about 8% at -10° angle of incidence to about 40% at 3° angle of incidence. The efficiency stays roughly constant at 40% from 3° to 6° incident angle and drops slowly with larger incident angles. This measured efficiency peak position is consistent with the simulated result referring to Fig. 2(c), but the maximum efficiency value is found (as expected) to be slightly lower than the simulated one. The main reason for this difference is fabrication error. The diameters of the three kinds of pillars were measured to be within ± 5 nm of the designed ones, which leads to the efficiency drop. Furthermore, as we can see from Fig. 3(c), the cross-section of the device shows that the three kinds of pillars have slightly different heights, and the tips of the pillars are not all flat, especially for the narrowest pillars. These fabrication defects are expected also to decrease the diffraction efficiency. In Appendix D we present a tolerance analysis which demonstrates that the efficiency curve in Fig. 4(b) can indeed be explained by the observed fabrication errors. Another possible discrepancy is that the laser is not perfectly Gaussian, so the power on the grating which we calculated by assuming a Gaussian profile for the illuminating beam may somewhat differ in reality. In conclusion, it is demonstrated that the metasurface shows an average diffraction efficiency of 25% for the NIR light within 20° full FOV from measurement.

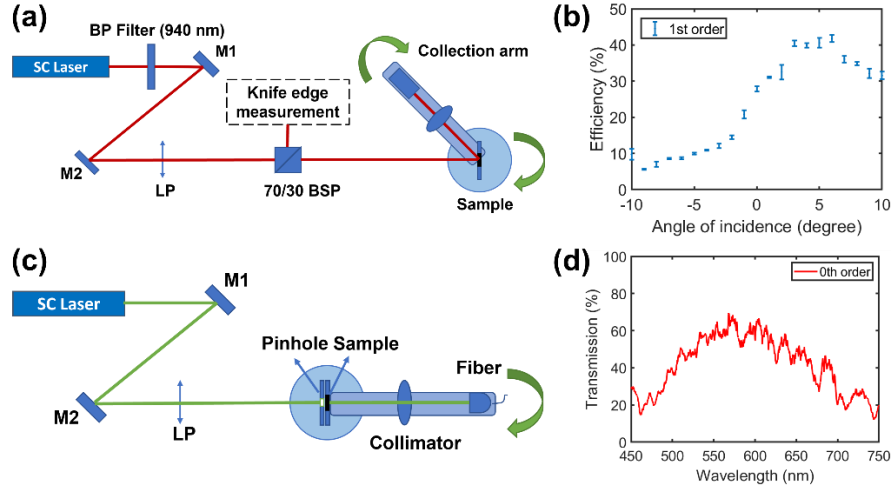

Figure 4. Metasurface measurements: (a) The experimental setup used to measure the diffraction efficiency: the super continuum (SC) laser passes through the band pass (BP) filter with a center wavelength of 940 nm and reflects from mirrors 1 and 2, then passes through the linear polarizer (LP). The 70/30 beam splitter (BSP) is used to transmit 70% of light for the grating efficiency measurement and reflect 30% of the light for the knife edge measurement. The 1$^{st}$ diffraction order on the reflection side of the sample is captured by the lens and detector on the collection arm. (b) Measured 1$^{st}$-order diffraction efficiency for the optimized metasurface as a function of the angle of incidence. (c) The experimental setup used to measure the transmission of the device: the SC laser reflects from mirrors 1 and 2, and then passes through the LP. Light transmits through the pinhole and illuminates the metasurface. A collimator collects the transmitted light and sends it to the fiber spectrometer. (d) The measured 0$^{th}$-order transmission spectrum of the fabricated device.

The transmission of the fabricated metasurface was measured on the setup illustrated in Fig. 4(c). The same white light laser source was used. Since the light source spot is larger than the metagrating area, a pinhole with a 0.5 mm diameter was placed in front of the fabricated device to let the laser beam underfill the sample (1 mm x 1 mm). This is to make sure all the light collected by the collimator and the fiber illuminates the metagrating area. The lens and photodiode on the collection arm were replaced with a collimator coupling the light into a fiber spectrometer (Ocean Optics HR2000CG-UV-NIR). We measured the spectrum of the beam collected by the fiber with and without the sample plate in the setup. The transmission was calculated by dividing the measured spectrum with the sample by the collected spectrum without the sample.

Figure 4(d) shows the measured 0$^{th}$ order transmission for the wavelength from 450 nm to 750 nm. The measured transmission is about 40% on average and it shows a peak of 60% at a wavelength of ~575 nm. The transmission drops to 20% at the two edges of the wavelength range. The measured transmission spectrum indicates that the fabricated metasurface is semi-transparent. The measured curve is also consistent with the line shape in Fig. 2(e) but with a 10% overall transmission drop. The difference can be explained by the same reasons as the diffraction efficiency mismatch. An additional reason could be that the deposited materials are not purely transparent due to defects during fabrication, which increases the absorption.

As mentioned at the end of the simulation section, the visible light diffracts when transmitting through the metasurface in simulation. In our measurement of transmission, we also noticed some rainbow-like artifacts. Song et al. minimized these artifacts by applying an absorptive material in their non-local metasurface [26]. However, this method is not that applicable if we want to achieve a high diffraction efficiency for redirecting near-IR light in the meantime. Minimizing these rainbow artifacts to increase the overall transmission in the visible while maintaining high diffraction efficiency can be explored in future work.

## 4. Conclusion

In this paper, we proposed and reported a hybrid VIS-NIR metasurface design that offers independent functionality across different wavelength bands. The dual function design is based on the combination of an aperiodic distributed Bragg reflector and dielectric meta tokens. We demonstrated by simulation, fabrication, and measurement that this metasurface device reflects NIR light into the desired angle while preserving transparency for the VIS light. We expect this dual-functional metasurface to open new functions for optical imaging and wearable devices, such as eye-tracking systems in AR displays.

**Appendix A**

This section compares the reflectance spectrum of the optimized aperiodic DBR substrate (Fig. 5(a)) and basic periodic DBR substrate (Fig. 5(b)). The red dashed line represents the 940 nm central wavelength and these two DBR substrates both have almost 100% reflectance at 940 nm. However, the optimized aperiodic DBR substrate has higher reflectance for the wavelength ranging from 450 nm to 750 nm (red semi-transparent area) compared to the periodic DBR substrate.

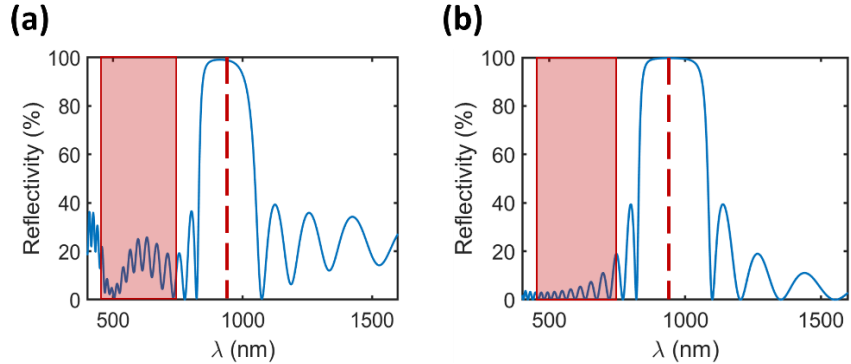

Figure. 5: DBR reflectance spectrum: (a). Reflectance spectrum for the optimized aperiodic DBR substrate; (b) Reflectance spectrum for the basic periodic DBR substrate.

**Appendix B**

This section shows the parameter sweep results of a single TiO2 pillar on top of the DBR substrate. We scanned the height of the pillar in a range from 100 nm to 650 nm and the radius of the pillar in a range from 0 to 200 nm. The resulting reflection amplitude and reflection phase are presented below.

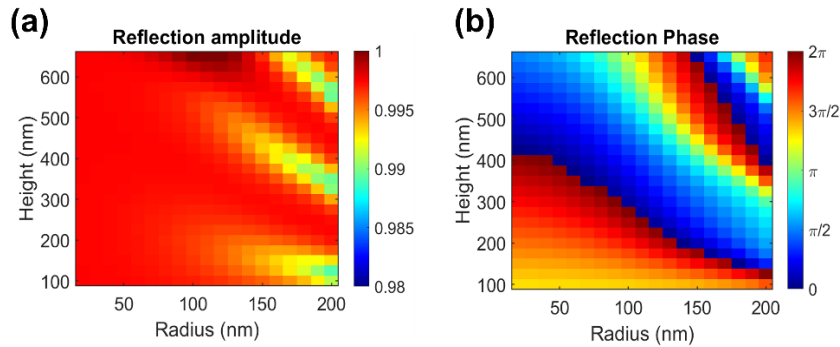

Figure. 6: Single TiO2 pillar sweep results: (a) Simulated reflection amplitude for various radii and heights of a single pillar; (b) Simulated reflection phase for various radii and heights of a single pillar.

**Appendix C**

This section shows the transmission angular bandwidth results of the optimized metasurfaces. We scanned the incident angle ranging from -70° to 70° and the wavelength from 450 nm to 750 nm to check the $0^{th}$ order transmission and the total transmission of the optimized metasurface as shown in Fig. 7 (a) and 7 (b), respectively. The difference between these two plots shows the diffraction of the visible light when transmitting through the metasurface. From Fig. 7 (a), the transmission angular bandwidth of this design is about 60°.

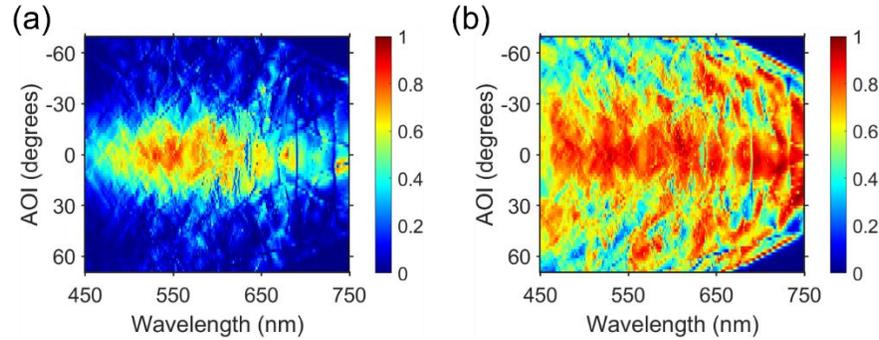

Figure. 7: Transmission angular bandwidth evaluation: (a) 0th order transmission efficiency map of nominal design for different incident angles and wavelengths; (b) Total transmission map of the nominal design for different incident angles and wavelengths.

**Appendix D**

This section shows simulation results modeling device performance degradation due to fabrication imperfections. We used the averaged measured results over the area of the device to determine the radii of the three "as-fabricated" pillars. The narrowest pillar has a different height compared to the other two pillars in the unit cell and the tips of all three pillars are rounded differently (based on Fig. 3(c)). The simulated efficiency result for this "as-fabricated" design is denoted by the orange line in Fig. 8(c), together with the simulated efficiency for the nominal design (blue line) and the measured efficiency for the fabricated sample (red dots). These results do confirm that the drop in performance can be explained by fabrication errors as in the real fabricated device the modeled fabrication errors are not uniformly present in each unit cell of the grating.

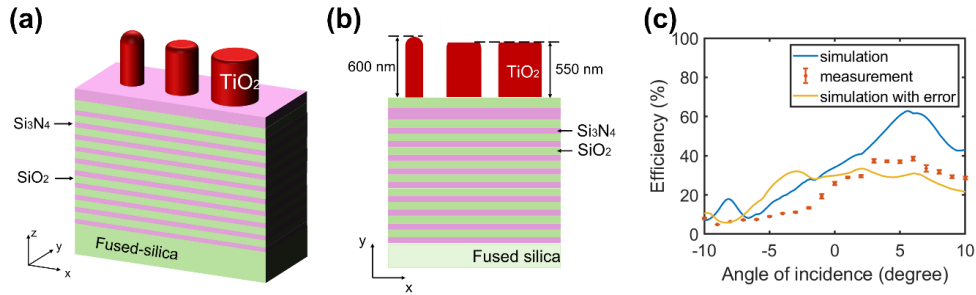

Figure 8: Fabrication artifacts simulation: (a) 3D image of the geometry for simulation of fabrication errors; the radius of the largest pillar is 185 nm, the radius of the middle pillars is 155 nm, and the radius of the smallest pillar is still 65 nm; (b) 2D image in the x-y plane of (a); (c) Comparison of the 1st order diffraction efficiency between the simulated result for the geometry in (a), the simulated result for nominal design and the measured result for fabricated sample.